\documentclass[traditabstract]{aa}
\usepackage{txfonts}
\usepackage{graphicx}

\newcommand{\msun}{\,{\rm M_\odot}}

\newcommand{\beq}{\begin{equation}}
\newcommand{\eeq}{\end{equation}}
\newcommand{\ba}{\begin{eqnarray}}
\newcommand{\ea}{\end{eqnarray}}
\def\spose#1{\hbox to 0pt{#1\hss}}
\newcommand{\lta}{\mathrel{\spose{\lower 3pt\hbox{$\mathchar"218$}}
      \raise 2.0pt\hbox{$\mathchar"13C$}}}
\newcommand{\gta}{\mathrel{\spose{\lower 3pt\hbox{$\mathchar"218$}}
      \raise 2.0pt\hbox{$\mathchar"13E$}}}
\def\simlt{\mathrel{\rlap{\lower 3pt\hbox{$\sim$}}\raise 2.0pt\hbox{$<$}}}
\def\simgt{\mathrel{\rlap{\lower 3pt\hbox{$\sim$}} \raise 2.0pt\hbox{$>$}}}

\begin{document}

\title{Limits on the high redshift growth of massive black holes}

\author{Ruben Salvaterra\inst{1} \and 
Francesco Haardt\inst{2,3}\and 
Marta Volonteri\inst{4} \and 
Alberto Moretti\inst{5}}

\institute{INAF, IASF Milano, via E. Bassini 15, I-20133 Milano, Italy \and
DiSAT, Universit\`a dell'Insubria, Via Valleggio 11, I-22100 Como, Italy \and
INFN, Sezione di Milano-Bicocca, Piazza delle Scienze 3, I-20123 Milano, Italy \and
IAP, 98bis Boulevard Arago, F-75014 Paris, France \and
INAF, Osservatorio Astronomico di Brera, Via Brera 28, I-20121 Milano, Italy}

\date{Received / Accepted}

\abstract{We place firm upper limits on the global accretion history of massive black holes at $z\gta5$ 
from the recently measured unresolved fraction of the cosmic X--ray background. 
The maximum allowed unresolved intensity observed at 1.5 keV implies 
a maximum accreted--mass density onto massive black holes  of
$\rho_{\rm acc} \lta 1.4 \times 10^4$ M$_\odot$Mpc$^{-3}$ for $z\gta 5$. Considering the contribution of lower--$z$ AGNs, 
the value reduces to $\rho_{\rm acc} \lta 0.66 \times 10^4$
M$_\odot$Mpc$^{-3}$. The tension between the need for the efficient and rapid accretion required by the observation of massive black holes already in place 
at $z\gta 7$ and the strict upper limit on the accreted mass derived 
from the X--ray background may indicate that black holes are rare in high redshift galaxies, or that accretion is 
efficient only for black holes hosted by rare galaxies. }


\maketitle

\section{Introduction}
While there is ample evidence that some supermassive black holes with masses 
exceeding $10^9$  M$_\odot$  formed as early as $z\gta  6$ 
(with a redshift record of $z =7.1$ reported by Mortlock et al. 2011; see also, e.g., Fan et al. 2001, Barth et al. 2003, 
Willott et al. 2009, Jiang et al. 2009), there is currently little or
no constraint on the evolution of the supermassive black--hole
population, as a whole, at the same redshifts. We have only been able
to probe the most exceptional quasars, which are powered by the most--massive black holes.

A powerful tool capable of globally constraining the nature of the
high redshift massive black--hole (MBH) population, at least of its active fraction, that is manifested as active galactic nuclei (AGNs), is the measure of the unresolved cosmic X--ray background (CXRB) (Dijkstra et al. 2004, Salvaterra et al. 2005, Salvaterra et al. 2007, McQuinn 2012).

{\it Chandra} deep observations have succeeded in resolving almost the
entire  (80-90\%) CXRB over its whole X-ray bandwidth (0.5--8 keV). The resolved fraction is almost $100\%$ at low energies, but decreases slightly, down to $\sim$ 85\%, at higher energies (see fig. 8 in Moretti et al. 2012, see also Moretti et al. 2003, Worsley et al. 2005). 
Cosmic X-ray background sources have been found to be mostly AGNs with some contribution
at soft energies ($<$ 2 keV) from galaxy clusters and starburst galaxies 
(Xue et al. 2011, Lehmer et al. 2012).
Most of the CXRB signal comes from sources located at $z\lta 2$, with only 
$\sim 1$\% being produced at $z\gta 4$ (Xue et al. 2011).
While in the hard band (2--10 keV) the residual unresolved fraction is 
commonly believed to be entirely due to the integrated
emission of undetected point sources, in the softer band (0.5--2 keV) most 
of the diffuse emission is due to thermal
radiation from the Galaxy and the local hot bubble (Kuntz \& Snowden 2000). 
A direct assessment of the unresolved fraction of the CXRB was  
performed by Hickox \& Marckevitch (2007) using {\it Chandra} deep field data. They 
found a small but statistically significant diffuse emission in the
1--2 keV band, but an emission consistent with zero at 
higher energies. However, the high {\it Chandra} instrument 
background, $\sim$25 times higher than the
unresolved  CXRB, makes this measure highly uncertain. 

Moretti et al. (2012) exploited the very low (compared to {\it Chandra}) 
instrument background of the {\it Swift} XRT to measure
the unresolved spectrum with the highest accuracy available today. 
This spectroscopic measure allowed the unresolved CXRB to be accurately probed with a much higher energy resolution. 
In particular, the constraint on the 1.5--2 keV band is a factor of three tighter than before.

In this letter, we take advantage of these new measurements of the unresolved fraction of the CXRB 
to put firm upper limits on the global accretion history of massive black holes at $z\gta5$. 
The aim of our approach is not to exclude a particular model but rather to 
highlight the existence of some tension between the need for efficient 
and rapid accretion required by the observation of supermassive black holes already in place 
at $z=7$ and the strict upper limit on the accreted mass of the whole 
high--$z$ MBH population imposed by the very tiny CXRB unresolved fraction.

We adopt a $(h,\Omega_m,\Omega_\Lambda)=(0.7,0.3,0.7)$ cosmology.

\section{Soltan's argument and the CXRB}

The so--called ``Soltan's argument" (Soltan 1982) translates the observed radiation emission of AGNs integrated over the 
cosmic history of the Universe into mass accreted onto the putative supermassive black-hole population. While the argument 
is usually expressed in terms of AGN luminosity functions, it is straightforward to apply it in the context of background radiation 
in a given observed band, which can easily be done as follows. 

Let us assume that the average AGN X-ray spectrum has a power--law form 
$L_\nu\propto E^{-\alpha}$, so that the comoving specific emissivity vs. redshift can be factorized as
\begin{equation}
j(E,z)=j_\star \left(\frac{E}{E_\star}\right)^{-\alpha}\, f(z).
\end{equation}
Here $f(z)$ is a function describing the redshift evolution of the emissivity. 
Neglecting absorption in the IGM, the contribution to the background at observed energy $E_0$ due to sources located at redshifts $z\geq {\bar z}$ is
\begin{equation}
J_{E_0}=\frac{1}{4\pi}\int_{\bar z}^\infty{dz\,\frac{dl}{dz}j(E,z)},
\end{equation}
where $E=E_0(1+z)$. If ${\bar z}\gg z_{m\Lambda}\simeq 0.33$ (the matter--$\Lambda$ equivalence redshift), we can 
neglect the cosmological constant energy density in the line element,
and by assuming $f(z)=(1+z)^{-\gamma}$ derive the normalization emissivity $j_\star$
\begin{equation}
j_\star=4\pi J_{E_0}\frac{H_0}{c}\Omega_m^{1/2}(E_0/E_\star)^{\alpha} (\alpha+\gamma+3/2) (1+{\bar z})^{\alpha+\gamma+3/2}.
\end{equation}

Now let us consider the standard Soltan's argument. The comoving mass density accreted onto MBHs within a given 
${\bar z}$ is 
\begin{equation}
\rho_{\rm acc}({\bar z})=\frac{(1-\epsilon)}{\epsilon c^2}\int_{{\bar z}}^{\infty}{dz\,\frac{dt}{dz}}\,\int_0^\infty{dE\,j(E,z)},
\end{equation}
where $\epsilon$ is the mass--radiation conversion efficiency. It is worth noting that at high redshifts $\rho_{\rm acc}$ could well be significantly lower than the total mass density locked in MBHs. We now relate the bolometric emissivity to the 
emissivity in a given energy band, $[E_m-E_M]$, by introducing a bolometric correction $f_X$ 
\begin{equation}
f_X\equiv \int_{E_m}^{E_M}{dE\,j(E,z)} \big/  \int_0^\infty{dE\,j(E,z)}.
\end{equation}
By substituting eq. 3 into eq. 4 and integrating, we finally obtain
\begin{eqnarray}
\rho_{\rm acc}({\bar z}) &=& 4\pi \frac{(1-\epsilon)}{\epsilon c^3}\frac{E_0J_{E_0}}{f_X (1-\alpha)}\frac{(\alpha+\gamma+3/2)}{(\gamma+3/2)} (1+{\bar z})^{\alpha}\times 
\nonumber \\
&& [(E_M/E_0)^{1-\alpha}-(E_m/E_0)^{1-\alpha}],
\end{eqnarray}
which is valid for $\alpha \neq 1$. The above formula allows us to estimate the maximum mass accreted 
onto MBHs within any given redshift interval, that contributes at any specified level to the observed background.

\section{Results}
We are interested in placing a firm upper limit on the mass accreted
onto MBHs prior to $z\gta 5$ by considering 
the unresolved  fraction of the CXRB.
The unresolved CXRB is well--described by a single power--law with a very hard photon-index ($\simeq 0.1\pm 0.7$) and a flux of $2.5^{+1.6}_{-1.3}\times 10^{-12}$ erg s$^{-1}$ cm$^{-2}$ deg$^{-2}$ in the 1.5--7 keV 
energy band. Comparing the measured unresolved CXRB to the AGN population model by Gilli et al. (2007), 
Moretti et al. (2012) find that most of the flux at $\simeq 1.5$ keV
can be accounted for by faint, $z\lta 5$ sources, but that their 
model falls short for $E\gta 3$ keV, suggesting that there is a larger population of Compton thick sources at moderate redshifts ($z\simeq 2$, see fig. 10 in Moretti et al. 2012). 
The maximum allowed (1--$\sigma$ error) intensity of the unresolved CXRB at 1.5 keV is $E_{1.5}J_{1.5}\simeq 0.47 
\times 10^{-12}$ erg s$^{-1}$ cm$^{-2}$ deg$^{-2}$. This value reduces to $ 0.21 \times 10^{-12}$ erg s$^{-1}$ cm$^{-2}$ deg$^{-2}$ when the contribution of the $z\lta 5$ faint sources modeled by Gilli et al. (2007) is taken into account. Eq. 6 derived in the previous section can be used to readily translate these limits into constraints on the total accreted mass density. To model the average Type I-Type II composite AGN spectrum, we adopt the spectral energy distribution (SED) proposed by Sazonov, Ostriker \& Sunyaev  (2004). In the 2--10 keV range, the spectrum is well--approximated by a power--law with $\alpha \simeq  0.25$, with a bolometric correction $f_X\sim 0.04$. Assuming $\epsilon=0.1$, for $E_0=1.5$ keV eq. 6 gives\footnote{We note that  $(\alpha+\gamma+3/2)/
(\gamma+3/2)\simeq 1$ for reasonable values of the cosmic evolution $\gamma$.}
\begin{equation}
\rho_{\rm acc}(z)\lta 3.4\times 10^4 \left( \frac{E_{1.5} J_{1.5}}{10^{-12}}\right) \left(\frac{0.04}{f_X}\right) 
\left( \frac{1+z}{7}\right)^{0.25} {\rm M}_\odot {\rm Mpc}^{-3}.
\end{equation}
In general, our mass density limit is valid for any X--ray emitting 
population that is not resolved above $z$ in the deepest X--ray surveys. For our purposes, the above limit is valid in particular for $z \gta 5$ as 
no AGNs above such a redshift limit has been identified in the 4 Msec CDF-S (Xue et al. 2011). 

In fig. 1, we plot the limits obtained by adopting the Moretti et al. (2012) results described above. The dark shaded region 
is excluded since it would imply an unresolved CXRB at 1.5 keV that is
higher than observed. The more stringent light-shaded region 
is obtained by subtracting the contribution of faint or absorbed
sources or both at $z\lta 5$. At $z=5$, the maximal possible accreted mass along the cosmic evolution of MBHs is 
$\rho_{\rm acc} \lta 1.4\times 10^4$ 
M$_\odot$Mpc$^{-3}$ ($\rho_{\rm acc}\lta 0.66 \times 10^4$ M$_\odot$Mpc$^{-3}$ subtracting the faint sources at lower redshifts).

\begin{figure}
\resizebox{\hsize}{!}{\includegraphics{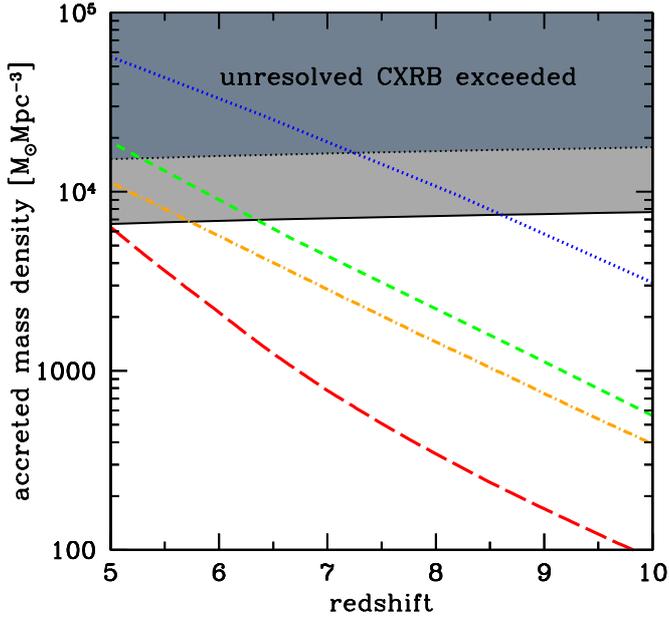}}
\caption{Limits on the density of accreted mass onto MBHs at $z\gta 5$ derived from the
  unresolved fraction of the CXRB observed at 1.5 keV. Dark shaded area
  refers to the maximum allowed CXRB intensity and light shaded area is
the limit once the contribution of lower-$z$ AGNs is taken into
account. For illustration, four curves show the accreted mass density
of models of formation and evolution of MBHs presented in Volonteri \&
Begelman (2010). These mass densities should be considered
qualitative, rather than quantitative, estimates, but they provide the
typical range found when assuming a fixed accretion rate for all MBHs
and self-regulated growth (see also Treister et al. 2011 for
additional examples). The dotted curve refers to  a model that induces
an early reionization, and the short--dashed curve to a model that barely reproduces the mass function of $z\sim 6$ quasars. These two curves assume massive MBH seeds and $f_{\rm Edd}=0.3$. The dot--dashed curve is analogous to the short--dashed model, but assumes the distribution of $f_{\rm Edd}$ given by Merloni \& Heinz (2008). The long--dashed curve is based on Population III remnants, and, while being consistent with the CXRB constraint,  is unsuccessful  in assembling $10^9$ M$_\odot$ MBHs by $z \simeq 6$.}
\label{fig1}
\end{figure}

The limits obtained above provide strong constraints on the
models of the formation and evolution of MBHs in the early
Universe. The accreted mass density is one of the most direct
predictions of semi-analytical models (see Volonteri 2010 for a
review) and can be directly compared with our constraints. 
We compare our limit on $\rho_{\rm acc}$ to simple  models (Volonteri
\& Begelman 2010), which assume
that all MBHs shine at a fixed  fraction, 30\%, of the Eddington
luminosity and that
accretion activity is major-merger driven and self-regulated by the host, assuming an unvarying relation to
the velocity dispersion at all redshifts. Gravitational wave recoil is neglected.  
In spite of the simple description of the
physical processes leading to the growth of the seeds 
(e.g., assumptions of a constant Eddington ratio for all objects
at all redshifts and of the local scaling relations being established in the high redshift Universe), 
these models are rather successful in reproducing the observed AGN
population. As an example, they reproduce the local MBH mass density, and the observed AGN bolometric
luminosity function at low-to-intermediate redshifts. 
We stress that the accreted mass density should be considered a
qualitative result, as Volonteri \& Begelman (2010) did not attempt to
model in detail the evolution of the MBH population (e.g., they
assumed a fixed accretion rate for all MBHs at all times, which is
clearly an oversimplification), but instead whished to estimate
whether a particular class of  seeds produced a physically reasonable
MBH population. Despite the qualitative nature of the theoretical
constraints, these models are very useful in putting the strict limit
on the accreted mass density implied by the CXRB in the context of MBH
evolution studies.

Models involving stellar-size Population III seeds may have difficulties in accounting 
for the existence of a MBH as massive as $M\gta 10^9$ M$_\odot$  at $z=7.1$, as  
observed by Mortlock et al. (2011) (see discussion in Petri, Ferrara, Salvaterra 2012). 
A model involving stellar-sized seeds is below our limits (lower long--dashed curve in Fig.~1), but fixing the accretion rate at 30\%  of the Eddington luminosity, it is unable to explain the presence of the 
population of $z\simeq6$ quasars (Willott et al. 2010). If we increased the accretion rate to 
100\% Eddington, to account for $z=6$ quasars, this model would overproduce the total unresolved CXRB (see Treister et al. 2011 for an example of this case).

Massive seeds, e.g., the so-called ``quasistars'' (see Volonteri 2010 and references therein), seem to be a more viable option to  
explain the observed population of high-z quasars.  This class of
models accounts for a population of MBHs  with $\sim 10^9\;\msun$  at
$z\simeq 6-7$. Volonteri \& Begelman (2010) provide two
observationally limited cases for the efficiency of the formation of
massive seeds.  The results in terms of $\rho_{\rm acc}$ are also
shown in fig. 1.  The high efficiency model (upper dotted curve)
results in a very early reionization of the IGM, while the low
efficiency one  (middle short-dashed curve) barely succeeds in
assembling enough MBHs as massive as  $10^9$ M$_\odot$ by $z \simeq
6$. Direct collapse models with fixed accretion rates all exceed our
limits. The model with the higher efficiency would  overproduce the
total unresolved CXRB at $z\simeq 7.2$,  while the low efficiency
model  is above the more stringent limit based on the subtraction of
lower-resdhift sources, for $z\gta 6.3$.  Similar results are found by
considering the MBH growth models of  Volonteri, Lodato \& Natarajan (2008), and Agarwal et al. (2012), among others. 

We stress that our result does not directly favour one seed model over another, but highlights the strong constraints on the average accretion rate of the high-z MBH population, as a whole. Assuming an initial mass density ($\rho_0$), and a mean Eddington ratio\footnote{Which can be considered a combination of accretion rate and duty cycle, see, e.g., Tanaka, Perna, \& Haiman 2012.} ($f_{\rm Edd}$), the accreted mass density is 
\begin{equation}
\rho_{\rm acc}(t)=\rho_0 \left[\exp \left(f_{\rm Edd}\frac{t}{\tau} \frac{1-\epsilon}{\epsilon} \right)-1 \right],
\end{equation}
 where $\tau=\sigma_T \,c/(4\pi \,G\,m_p)= 0.44$ Gyrs ($c$ is the speed of light, $\sigma_T$ is the Thomson cross--section, and $m_p$ is the proton rest mass). This approximation requires that all seeds have similar masses and form roughly at the same time. Additionally, such a generic argument does not take into account any self-regulation or feedback effect that limits the MBH growth. Keeping these caveats in mind, for plausible values of $\rho_0\simeq 10-1000$ M$_\odot$ Mpc$^{-3}$, the average $f_{\rm Edd}$ must be less than 0.1--0.3 at $z\gta 5$.  This is in line with lower--redshift results 
that the distribution of Eddington rates of  $z=2-4$ luminous quasars is dominated by 
sub--Eddington sources (Kelly et al. 2010). With such an average $f_{\rm Edd}$, the $M=2\times 10^9$ M$_\odot$ black--hole observed by Mortlock et al. (2011) at $z=7.1$  would require a seed black-hole of mass exceeding $2\times 10^7$ M$_\odot$. 
We conclude that the most massive MBHs at very high redshift cannot accrete at the average Eddington ratio 
(see also Trakhtenbrot et al. 2011). Models require rates close to Eddington to 
explain the high--mass end of the mass function of quasar-powering MBHs (Natarajan \& Volonteri 2011).

\section{Possible solutions}

Massive black--hole growth models seem to predict $\rho_{\rm acc}$
above our observational limit. One may argue that the bulk of the
accreted mass is missing in the small volume sampled by the CDFs. A
simple estimate implies that objects rarer than $\sim 10^{-6}$
Mpc$^{-3}$ are not present in the field. However, even assuming that
all of these BHs were accreted up to $\sim 10^8$ M$_\odot$, the resulting $\rho_{\rm acc}$ would be much lower than our limit, further increasing the disagreement with evolutionary models. 

On the theoretical side, models tested here are based on simple assumptions. 
As an example, $\rho_{\rm acc}$ is expected to be reduced by
gravitational wave recoil, which we however neglect. 
We did check that in the considered models the effect on $\rho_{\rm
  acc}$ is at most $\simeq 20$\%, hence that the inclusion of the
recoil does not 
represent a viable solution. 
A stronger impact, but still insufficient, is obtained by considering  a non--fixed accretion rate. 
A model starting from massive seeds coupled to the empirical distribution of Eddington ratios derived by Merloni \& Heinz (2008), shown in fig. 1 as a dot--dashed line, exceeds our limit at $z\lta 6$. 

The most promising solution relies on the possibility that the most massive black holes are able 
to maintain a high $f_{\rm Edd}$ during their cosmic history, while
lighter ones accrete at a much lower rate. 
Therefore, accretion must strongly depend on either the MBH mass (or
most likely the host mass as in `selective accretion', Volonteri \&
Stark 2011) or environment (di Matteo et al. 2012), or must have a
low--mass cut-off (``global warming", Tanaka et al. 2012). This is in
line with the observational evidence that high redshift quasars seem to be powered by 
MBHs that are `over-massive' for a fixed galaxy property with respect
to their counterparts at $z=0$ 
(e.g., Wang et al. 2010).  Willott et al. (2010) instead find that
either many massive galaxies at $z=6$ do not have MBHs, or that these MBHs are less massive than expected assuming that MBHs are roughly 1/1000 of the 
host stellar mass. This 
suggests overall that while some (most--likely the most massive) MBHs can grow above today's correlations,
most of them should be less massive than expected from local relations (cf.  Volonteri \& Stark 2011).

\section{Conclusions}

We have placed firm upper limits on the global accretion history of
MBH at $z\gta 5$ by taking advantage of  the 
measurement of the unresolved fraction of the CXRB provided  by Moretti et al. (2012). 
The maximum allowed unresolved CXRB intensity observed at 1.5 keV implies 
a maximum accreted--mass density onto MBH at $z\gta 5$ of 
$\rho_{\rm acc} \lta 1.4 \times 10^4$ M$_\odot$Mpc$^{-3}$. Considering the contribution of lower--$z$ 
AGNs\footnote{Adopting the speculative model of Moretti et al. (2012)
  which is able to explain the very hard shape of the 
unresolved 1.5--7 keV CXRB, the limit on $\rho_{\rm acc}$ further decreases by a factor of $\simeq 2$.}
(Gilli et al. 2007), this limit reduces to $\rho_{\rm acc} \lta 0.66 \times 10^4$ M$_\odot$Mpc$^{-3}$. This value translates 
into $\lta 1$ HI ionizing photon per baryon produced by accretion onto MBHs at $z\simeq 6$, confirming the common 
wisdom that hydrogen reionization is driven by stellar--like sources (cf. Haardt \& Madau 2012).   

It is important to stress that this calculation is a strict upper limit to the accreted mass onto MBHs at $z\gta 5$, as 
most of the unresolved CXRB could well be attributed to faint sources
at lower redshift that are not accounted for by the 
Gilli et al. model. That the spectrum of the unresolved CXRB determined 
by Moretti et al. (2012) is very hard, alone seems to exclude there be
a significant contribution 
from AGNs at $z\gta 5$. For such a population, the Compton reflection roll--over  
(rest--frame energy $\simeq 30$ keV) would fall within the observed energy band, resulting in an emission  
much softer than the observed unresolved CXRB. Therefore, a significant contribution from high--$z$ sources 
would result in an even harder spectrum of the still unaccounted fraction of the CXRB, possibly at odds with 
current population--synthesis models. 

A possibly stricter upper limit on $\rho_{\rm acc}$ at $z\simeq 6.5$ can be obtained by the stacking 
analysis of the X--ray emission of {\it i}--dropouts selected by
Bouwens et al. (2006) in the CDF--S. In contrast to Treister et
al. (2011), Willott (2011), Fiore et al. (2012), and Cowie et
al. (2012) did not find any evidence of X--ray emission. The flux limit in the 
observed hard X--ray band derived from the stacking analysis corresponds to a MBH mass density
of $\rho_{\rm acc} \lta 0.4 \times 10^4$ M$_\odot$Mpc$^{-3}$. A much
stronger, about ten times lower, upper limit is obtained from the flux
limits obtained using the more
sensitive soft X--ray band, though here absorption can play a decisive role.
These
limits are tighter  than ours at the same redshift, although we believe
that our results are less subject to biases and assumptions than those
derived from the stacking analysis. In
particular: i) the stacking analysis relies on the corrections for
incompleteness, photometric redshift measurements, and dust absorption of the Bouwens et al. sample; ii)
the flux limit of the {\it i}--dropouts implicitly introduces a lower
limit on the MBH mass probed (that we estimate to be a few times $10^6$
M$_\odot$); and iii) the stacking analysis
results strictly refer only to the AGN activity in a narrow redshift
range. 
The limits we place on $\rho_{\rm acc}$ are instead unaffected by any of these effects since the background intensity directly measures 
the time--integrated accreted mass.

We investigate how much our assumptions on the emission properties of 
high--$z$ MBHs affect our results. The adoption of a much flatter SED, 
e.g. with $\alpha=0.9$, will result in limits that are $\simeq 1.8$ less stringent 
than those shown in fig. 1. On the other hand,  as 
shown by Marconi et al. (2004), the fraction of light emitted by AGNs in the 
rest-frame 2--10 keV band increases with decreasing bolometric 
luminosities, so that $f_X=0.04$ can 
be considered a conservative choice. We note that at $z\gta 5$ the observed 1.5 keV photons has been emitted   
at energy $\gta 9$ keV, where absorption can not be 
very strong even for high intrinsic column--densities. Thus, even assuming that 
high--$z$ AGNs are heavily obscured, our limits cannot increase by much. 
Finally, our adopted value of the accretion 
efficiency, $\epsilon=0.1$, is in--between the range of allowed values. 
For Schwartzschild BHs ($\epsilon=0.057$), the limits would be 1.7 higher, 
while for maximal rotating BHs ($\epsilon=0.42$) they are a factor of six lower. 

 The bottom line of our analysis is that there is some tension between the need for efficient 
and rapid accretion required by the observation of SMBHs already in place 
at $z=7.1$ (Mortlock et al. 2011), and the strict upper limit on the accreted mass derived 
from the CXRB. Therefore, accretion must be very efficient for the
most massive BHs during their lifetime, 
but sub--Eddington for most of the AGN population.

\begin{acknowledgements}We thank F. Fiore for helpful discussions and comments, and the anonymous referee 
for her/his useful suggestions. \end{acknowledgements}

\end{document}